\documentclass[12pt]{article}
\usepackage{a4}
\usepackage{epsfig,url}
\usepackage{lineno}

\newcommand\vr{\vec r}
\begin{document}
\title{Closed-form expressions for the magnetic field of permanent
magnets in three dimensions}
\author{V. Ziemann, Uppsala University}
\date{\today}
\maketitle
\begin{abstract}\noindent
  We derive a closed-form expression of the magnetic field of a
  finite-size current sheet and use it to calculate the field
  of permanent magnets, which are modeled through their surface current
  densities. We illustrate the method by determining the multipoles and
  the effective length due to fringe fields of a finite-length dipole
  constructed of magnetic cubes.
\end{abstract}
%
%
\section{Introduction}
Today permanent magnets are widely used in the construction of efficient electrical motors,
for example those used in battery-driven automobiles, but also for scientific applications,
such as ion traps to store and analyze rare atoms, or for undulators to generate synchrotron
radiation, and for multipole magnets to guide and confine charged-particle beams. These
magnets do not need to be powered and are therefore very energy efficient, which makes
them attractive regarding today's focus on sustainability. In the 1970s Klaus Halbach
showed~\cite{KH1} how to design multipole magnets and undulators using paper and pencil
only. His approach made use of variables in the complex plane and is essentially limited
to two spatial dimensions. In order to analyze three-dimensional features we typically
use finite-element codes~\cite{FEM}. In this report, we offer a complementary approach
based on solving Biot-Savart's equation for a finite-size rectangular current sheet and modeling
the magnets as an assembly of such sheets. In iron-free geometries this allows us to calculate the three
components of the magnetic field $\vec B(z)$ at any point in space as the superposition
of the contributions from the current-carrying sheets. Since all integrations are done
analytically this results in expressions that make calculating the fields very rapid.
Moreover, the field values are exact, no approximations due to finite meshes are involved.
\par
A key feature of permanent magnet material is that the relative magnetic permeability
$\mu_r$ is very close to unity, which makes them transparent to other, externally generated
fields. As a consequence, the field at a given point $\vec z$ is the superposition of
the fields due to all current sources. And here the second key feature enters; the magnetic
field of permanent magnets can be calculated from the equivalent surface current density
$I'=dI/ds =B_r/\mu_0$ that is proportional to the remanent magnetic field $B_r$ of the
permanent magnet material. Here $\mu_0=4\pi\times 10^{-7}\,$A/Tm is the permeability of
free space. Since $B_r$ is on the order of one Tesla, the corresponding current densities
are enormous.
\par
These observations now guide us through the remainder of this report. We first calculate
the field of a straight current-carrying filament, followed by spreading the filament
laterally to obtain a current sheet and calculating the field it generates. In the
subsequent section we briefly introduce software, based on MATLAB, to assemble magnets
that are composed of these sheets and calculate their fields. We then use the software
to illustrate the field of a single sheet, a magnet cube assembled of four sheets, and
a dipole magnet, made of four cubes, whose field quality and fringe fields we analyze.
Finally, we consider solenoid magnets which help us to validate the calculations
and the software.
%
\section{Field of a filament}
The magnetic field at point $\vr_2$ due to a filament that carries a current $I$, starts
at position $\vr_a$, and ends at $\vr_b$ is given by Biot-Savart's law
\begin{equation}
\vec B(\vr_2) = \frac{\mu_0 I}{4\pi}\int \frac{d\vr_1\times (\vr_2-\vr_1)}{\vert\vr_2-\vr_1\vert^3}
=  \frac{\mu_0 I}{4\pi} \int_0^1 
\frac{[\vr_b-\vr_a]\times[\vr_2-t\vr_b-(1-t)\vr_a]}{\vert\vr_2-t\vr_b-(1-t)\vr_a)\vert^3} dt\ ,
\end{equation}
where we parameterize a point on the filament $\vr_1$ by $\vr_1=(1-t)\vr_a+t\vr_b$ with $0<t<1$.
First we simplify the numerator in the last equality and find
\begin{equation}
(\vr_b-\vr_a)\times(\vr_2-t\vr_b-(1-t)\vr_a) = (\vr_b-\vr_a)\times(\vr_2-\vr_a)
\end{equation}
whereas for the denominator we obtain
\begin{eqnarray}
\vert\vr_2-t\vr_b-(1-t)\vr_a)\vert^2&=& (\vr_2-t\vr_b-(1-t)\vr_a)\cdot (\vr_2-t\vr_b-(1-t)\vr_a)
\nonumber\\
&=& (\vr_2-\vr_a)^2 - 2t(\vr_2-\vr_a)\cdot(\vr_b-\vr_a) + t^2(\vr_b-\vr_a)^2\nonumber\\
&=& A-2Bt+Ct^2
\end{eqnarray}
with 
\begin{equation}\label{eq:ABC}
A= (\vr_2-\vr_a)^2,\quad B= (\vr_2-\vr_a)\cdot(\vr_b-\vr_a), \quad\,\mathrm{and}\quad
C=(\vr_b-\vr_a)^2\ .
\end{equation}
Note that these parameters have simple geometric interpretations: $A$ describes the squared
distance $d_{P}$ between the observation point $P$ at $\vr_2$ and the starting point of the
filament $\vr_a$. The variable $B$ is the product of $d_P$, the length $L$ of the filament,
and the cosine of the enclosed angle, whereas $C$ is the length of the filament squared.
Inserting these expressions into the equation for the magnetic field $\vec B$ we get
\begin{equation}
\vec B(\vr_2)=\frac{\mu_0 I}{4\pi}(\vr_b-\vr_a)\times(\vr_2-\vr_a)\int_0^1\frac{dt}{(A-2Bt+Ct^2)^{3/2}}\ .
\end{equation}
This integral can be found in integral tables~\cite{GR} and is given by
\begin{equation}
\int \frac{dt}{(A-2Bt+Ct^2)^{3/2}} = \frac{Ct-B}{(AC-B^2)\sqrt{A-2Bt+Ct^2}}
\end{equation}
such that we write $\vec B(\vr_2)$ as
\begin{equation}\label{eq:Bfil}
\vec B(\vr_2)=\frac{\mu_0 I}{4\pi}\frac{(\vr_b-\vr_a)\times(\vr_2-\vr_a)}{AC-B^2}
\left[\frac{C-B}{\sqrt{A-2B+C}}+\frac{B}{\sqrt{A}}\right]
\end{equation}
where $A,B$ and $C$ are given in Equation~\ref{eq:ABC}.
Equation~\ref{eq:Bfil} is already quite useful to determine the fields of systems
containing a finite number of current-carrying wires. We previously used it to design 
earth-field compensation coils~\cite{VZEMC}.
\section{Field of a finite current sheet}
%
\begin{figure}[tb]
\begin{center}
\includegraphics[width=0.6\textwidth]{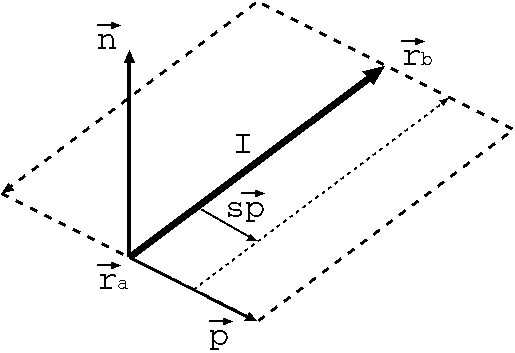}
\end{center}
\caption{\label{fig:cs}Current sheet.}
\end{figure}
In order to calculate the field of a sheet, we first consider the field that is generated
by a filament that is laterally displaced by the vector
\begin{equation}
  \vec p = \alpha (\vr_b-\vr_a)\times\vec n\ ,
\end{equation}
where $\vec n$ is the vector normal to the sheet and $\alpha$ is the aspect ratio of
the sheet, such that $w=\alpha\vert\vr_b-\vr_a\vert=\alpha\sqrt{C}$ is the width of
the sheet. The construction is illustrated in Figure~\ref{fig:cs}.
Note that $\vec p$ is perpendicular to the filament; the scalar product of
$\vr_b-\vr_a$ and $\vec p$ is zero. This construction permits us to parameterize the
parallel-displaced filament by its end points $\vr_a+s\vec p$ and $\vr_b+s\vec p$,
where $-1/2\leq s\leq 1/2$ describes by how far the filament is displaced, as shown
in Figure~\ref{fig:cs} by the thin dotted line that is parallel to $\vr_b-\vr_a$.
The field from the displaced filament is then given by replacing the parameters
$A, B$ and $C$ by their $s$-dependent counterparts given by
\begin{eqnarray}
  A(s)&=&  (\vr_2-\vr_a-s\vec p)^2\nonumber\\
  B(s)&=& (\vr_2-\vr_a-s\vec p)\cdot(\vr_b-\vr_a)\\
  C(s)&=& (\vr_b+s\vec p-\vr_a-s\vec p)^2 = C\ .\nonumber
\end{eqnarray}
Moreover, the cross-product before the integral becomes $(\vr_b-\vr_a)\times(\vr_2-\vr_a-s\vec p)$.
We now ``spread out'' the current $I=I'w$ evenly across the width $w=\alpha\sqrt{C}$
of the sheet with constant current density $I'$ and sum up the contributions of all
displaced filaments, which corresponds to integrating over $s$. We also note that the
current density $I'$ is related to the remanent magnetic field $B_r$ through $\mu I'=B_r$.
Combining these concepts we find that the the field at point $\vr_2$ is given by the
following integral
\begin{eqnarray}\label{eq:intds}
  \vec B(\vr_2)
  &=&\frac{B_r\alpha \sqrt{C}}{4\pi}\int_{-1/2}^{1/2}
      \frac{(\vr_b-\vr_a)\times(\vr_2-\vr_a-s\vec p)}{A(s)C-B(s)^2}\\
  &&\qquad\qquad\quad\times
     \left[\frac{C-B(s)}{\sqrt{A(s)-2B(s)+C}}+\frac{B(s)}{\sqrt{A(s)}}\right] ds
  \nonumber
\end{eqnarray}
where $A(s)$ and $B(s)$ are at most quadratic functions of $s$. The evaluation of the
integral, which is rather lengthy and therefore deferred to Appendix~\ref{sec:appA},
leads to
\begin{eqnarray}\label{eq:Br2}
 \vec B(\vr_2)
  &=&\frac{B_r\alpha C}{4\pi}
     \left[(C-B)J_1 + B J_3\right] (\vr_b-\vr_a)\times(\vr_2-\vr_a)\\
  &&\qquad +\frac{B_r\alpha C^2}{4\pi} \left[(C-B) J_2 + B J_4\right]\vec n\ ,\nonumber
\end{eqnarray}
where $J_1$ to $J_4$ are defined in Appendix~\ref{sec:appA} in Equations~\ref{eq:JJ}
to~\ref{eq:I2}. We point out that $\alpha C$ is the area of the sheet, such that
$B_r\alpha C$ describes the integrated ``strength'' of the sheet to excited magnetic
fields.
\section{Software implementation}
\label{sec:software}
Equation~\ref{eq:Br2} describes the magnetic field vector $\vec B(\vr_2)$ that is caused
by one sheet that is characterized by $\vr_a, \vr_b, \vec n, \alpha, B_r$ in a coordinate-free
way and only requires function evaluations; no integrations are necessary which makes the
calculation of $\vec B(\vr_2)$ very fast. In order to calculate the field at a given
point $\vr_2$ due to several sheets we only have to sum over the contributions from each
of the sheets.
\par
We therefore represent each sheet by a one-dimensional array that contains the three
spatial components of  $\vr_a, \vr_b,\vec n$ as well as $\alpha$ and $B_r$ for that sheet;
thus there are eleven numbers to store for each sheet. The entire geometry is then given
as a collection of, say $M$, such sheets that is stored in an array of dimension $M\times 11$.
\par
In order to place the sheets within the geometry, we prepare functions to translate
sheets by a vector $\vec a$; the function simply adds $\vec a$ to $\vr_a$ and $\vr_b$
in the array with the sheets. Likewise, three functions to rotate a sheet by an angle
$\theta$ around each of the three axes, just multiply $\vr_a, \vr_b$, and $\vec n$
in a sheet by the appropriate rotation matrix. Using these functions, it is easy to
create any geometry, very similar to the way that three-dimensional modeling software
composes complex geometries from geometric primitives, often triangles, but also from
cubes or spheres.
\par
Finally functions to calculate the field at a point or along a path just loop over
the sheets and add their contributions. We implemented these functions in MATLAB,
briefly explain the functions in Appendix~\ref{sec:appB}, and illustrate their use
with a number of simple examples in the next section.
\section{Examples}
%
\begin{figure}[tb]
\begin{center}
\includegraphics[width=0.47\textwidth]{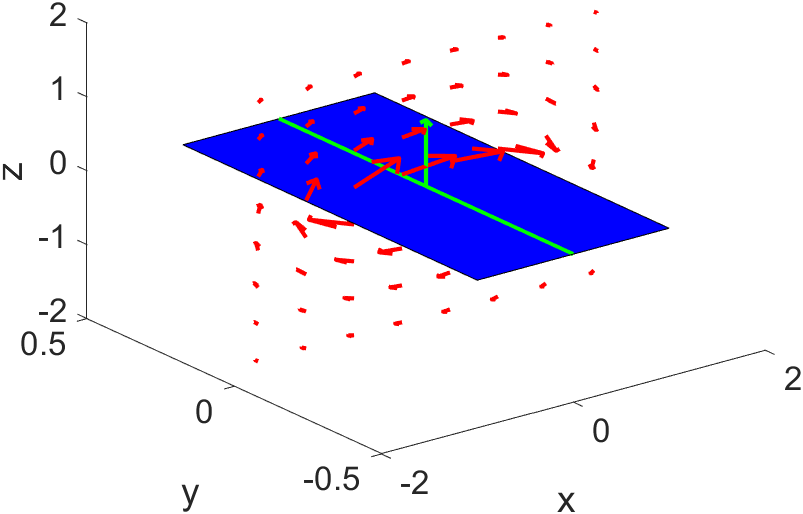}
\includegraphics[width=0.47\textwidth]{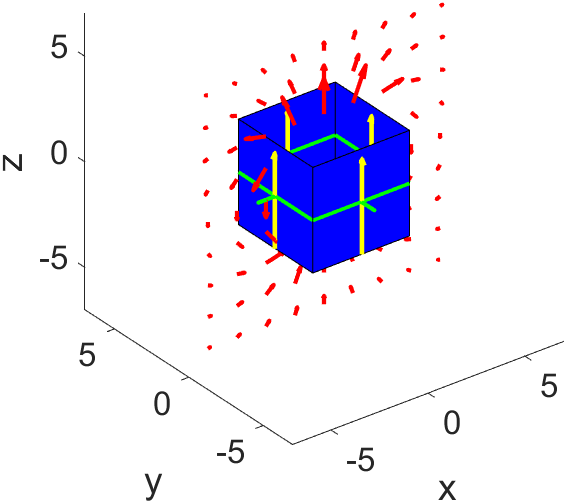}
\end{center}
\caption{\label{fig:one}Field pattern of a current sheet (left) and a
  four sheets assembled to generate the field of a permanent magnet
  cube (right).}
\end{figure}
As a first example we consider the field of a single sheet, which is shown on the left-hand
image in Figure~\ref{fig:one}. The sheet is shown as the blue rectangle and the current
as the green line stretching in the center of the sheet. The vertical green arrow indicates
the normal vector $\vec n$ of the sheet. We observe that the magnetic field vector $\vec B$
curls around the sheet, perpendicular to the current's direction, pointing to the right
above and to the left below the sheet. Making the sheets very narrow causes the field
to approach a circular form, just as expected from Ampere's law.
\par
In the next step we assemble four quadratic current sheets to represent a permanent magnet
cube, which is shown on the right-hand image in Figure~\ref{fig:one}. We see that the
current, again shown as a green line, flows around the surface of the cube. We use the
convention that the normal vectors, again shown as green arrows, point towards the outside
of the magnet. The easy axis of the magnet is indicated by the yellow arrow on the sheets,
that is oriented perpendicular to the current flow and points upwards. The magnetic field
created by the four sheets representing the permanent magnet cube is illustrated by the
red arrows. We observe that the field points upwards at the top of the cube, then turns over
such that it points down on the outside of the cube, then turns over again to point
upwards below the cube. We point out that the four current sheets form a square solenoid
and the field therefore corresponds to that geometry.
\par
\begin{figure}[p]
\begin{center}
\includegraphics[width=0.49\textwidth]{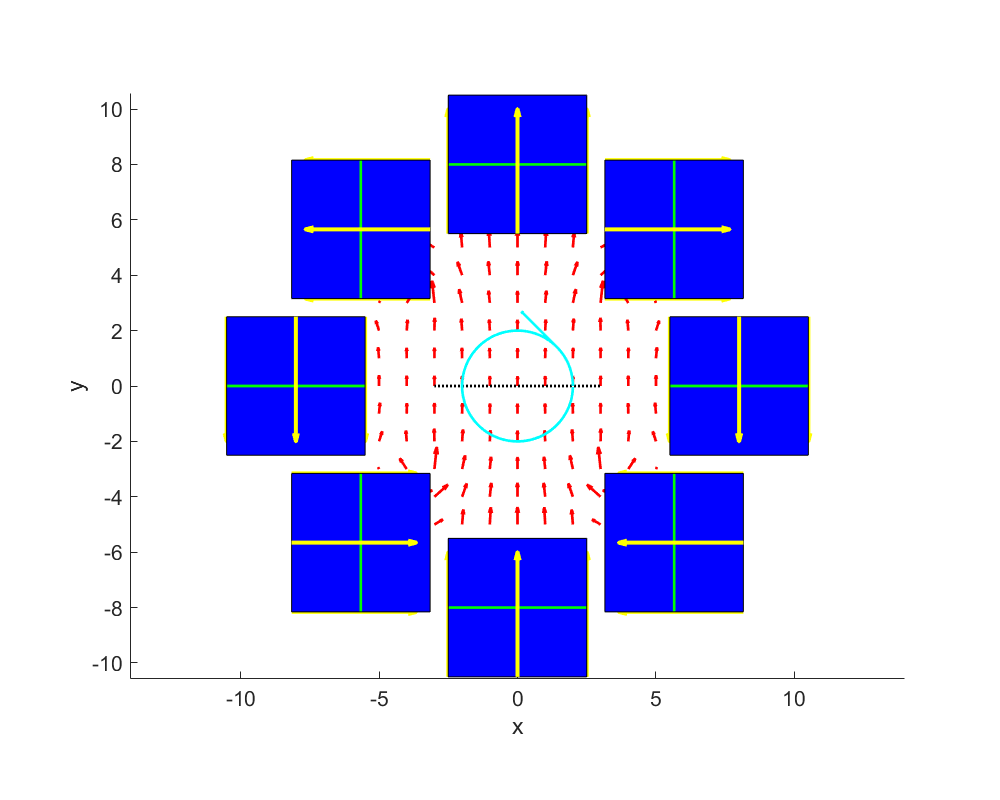}
\includegraphics[width=0.49\textwidth]{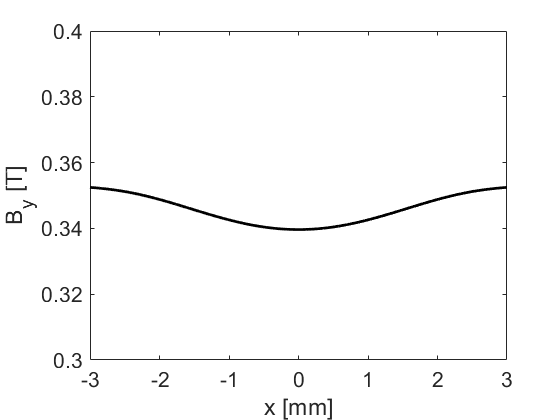}
\includegraphics[width=0.49\textwidth]{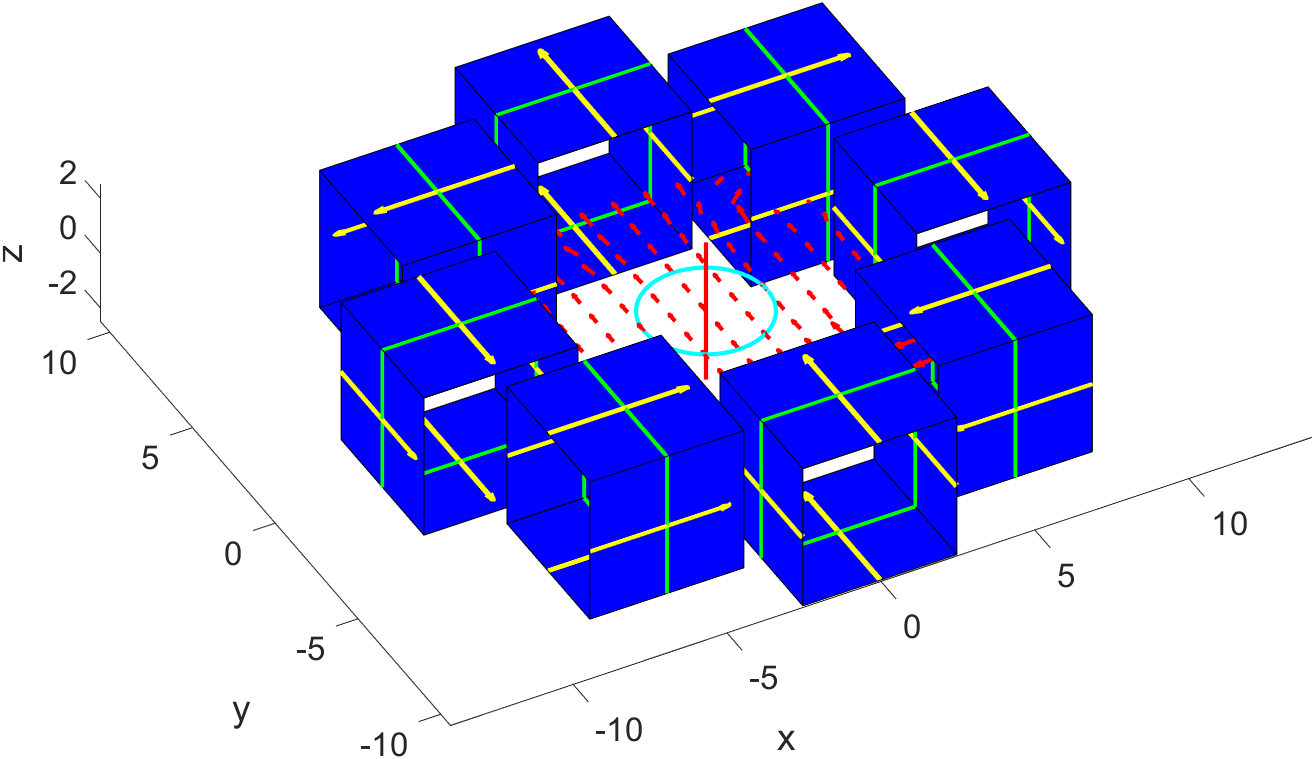}
\includegraphics[width=0.49\textwidth]{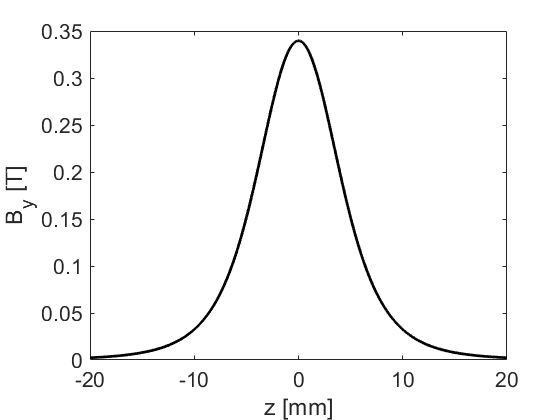}
\includegraphics[width=0.49\textwidth]{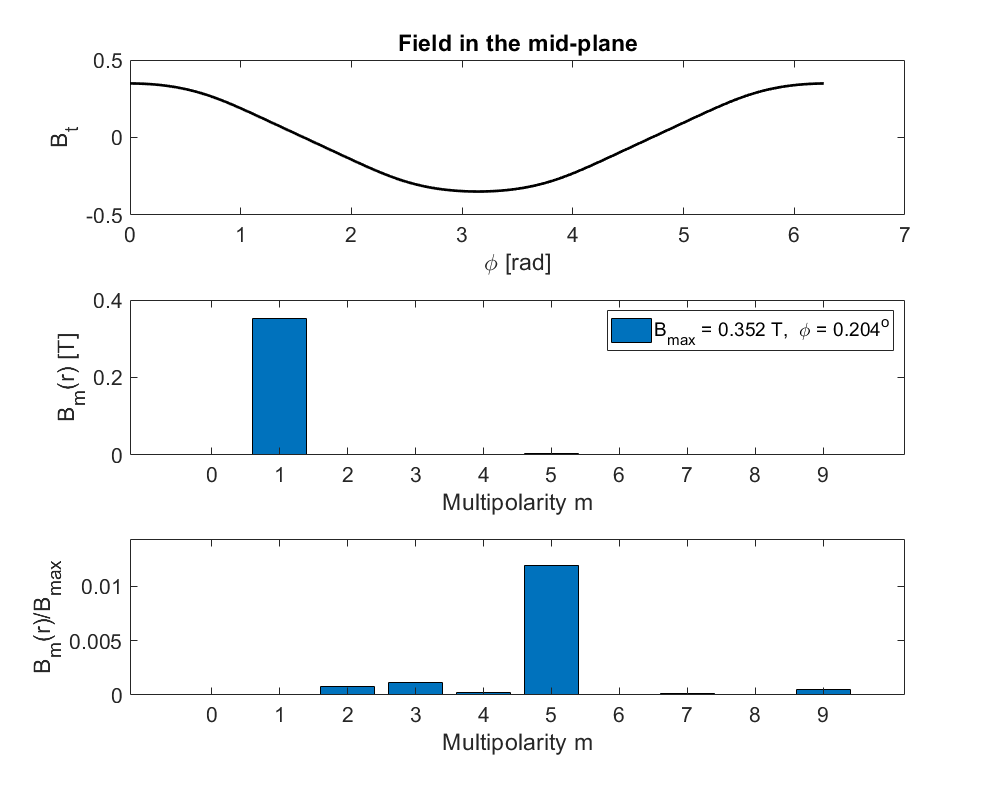}
\includegraphics[width=0.49\textwidth]{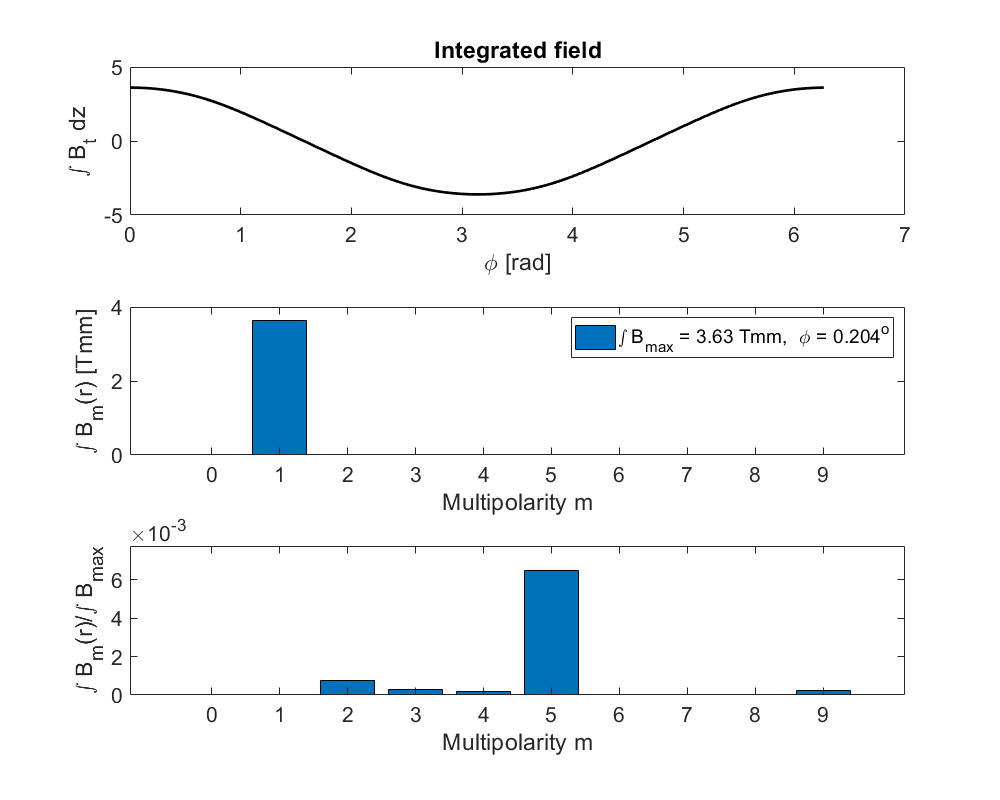}
\end{center}
\caption{\label{fig:hdip} The geometry of a permanent magnet solenoid (top left) and
  the vertical field in the midplane in the $x$-direction (top right). A 3D rendition
  of the magnet (middle left) and the field along the $z$-axis (middle right). The
  bottom left shows the field along the circle shown in cyan and the multipole
  coefficients (absolute and relative) in the midplane. The bottom right shows the
  corresponding integrated values.}
\end{figure}
We reach the next level of complexity by assembling eight 5\,mm-cubes with $B_r=1.47\,$T
to form an approximation
of a Halbach dipole, discussed in Appendix~A.3 in~\cite{VZAP}. The top left image in
Figure~\ref{fig:hdip} shows the view onto the assembly from the top. The easy axis
of the magnets, shown by the yellow arrows, rotates twice when moving from one magnet
to the next; in every fourth of the eight
magnets it points in the same direction. In the center of the assembly, the magnetic
field points upwards, as is expected for a dipole. The plot on the top right shows the
vertical field component $B_y$ along the black dotted line shown on the left-hand image.
We observe that the field is on the order of 0.35\,T and has a small dip in the center.
The image on the left-hand side in the middle row shows a perspective view of the same
geometry which makes the three-dimensional character obvious. The vertical red line
in the center of the assembly indicates the direction of the $z$-axis and the plot
on its right shows $B_y$ along this direction, but extending over $\pm 20\,$mm.
We see that the peak field of about 0.35\,T appears in the center of the magnet, but
the fringe field extends significantly outside the magnet, which only has a a height
of $\pm 2.5\,$ mm.
\par
We determine the multipolarity of the assembly at the radius $r$ shown by the cyan
circle on he top-left image. A tangent to the circle, illustrated by the cyan straight
line, is also shown. In order to  determine the multipoles we calculate the scalar
product of the tangent vector with the magnetic field at every point on the circle.
This yields the tangential field component $B_t$ which we display in the top
panel of the image at the bottom left plot as a function of the azimuthal angle $\phi$.
The multipolarity, shown on the middle panel then follows from a Fourier-transformation,
as discussed in Appendix~\ref{sec:appC}. We find that mainly the dipole component with
multipolarity $m=1$ is present and that its amplitude is $B_{max}=0.353\,$T. Moreover, from the
real and imaginary part of the transform, we determine the roll angle of the magnet,
which here turns out to be close to zero. On the bottom panel we show the amplitudes of
the higher multipoles, normalized to $B_{max}$. We find the decapole component ($m=5$)
with a relative amplitude on the order of a percent to give the dominant contribution.
Finally, we calculate the integral of $B_t$
along a line in the range $z=\pm 20\,$mm with $x$ and $y$-positions determined by the
cyan circle and display it on the upper panel in the image on the bottom-right of
Figure~\ref{fig:hdip}. Again, mostly the dipole component with $m=1$ is present.
Moreover the field integral turns out to be $3.63\,$Tmm, which allows us to determine
the effective length of the magnet to be about 10.3\,mm, a value that significantly
exceeds the physical length of the magnet, which is only 5\,mm. Again the decapole
component with a relative magnitude of about $6\times10^{-3}$ gives the dominant
contribution to the integrated field.
\par
\begin{figure}[tb]
\begin{center}
\includegraphics[width=0.47\textwidth]{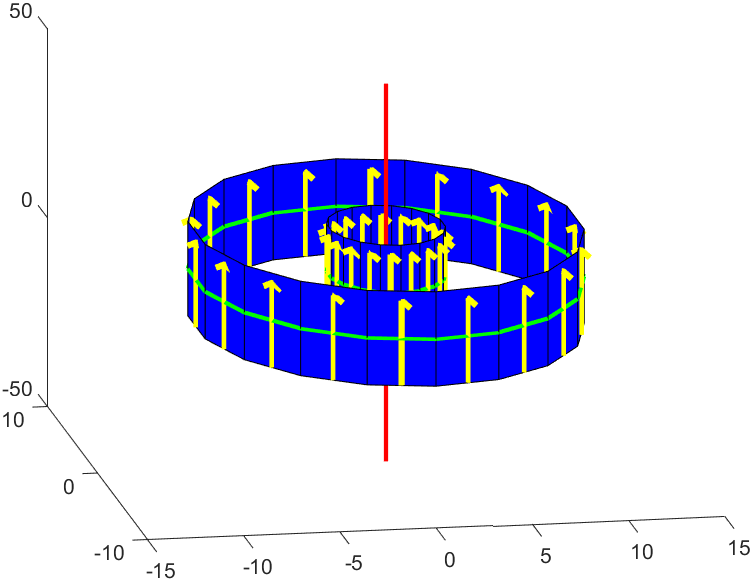}
\includegraphics[width=0.47\textwidth]{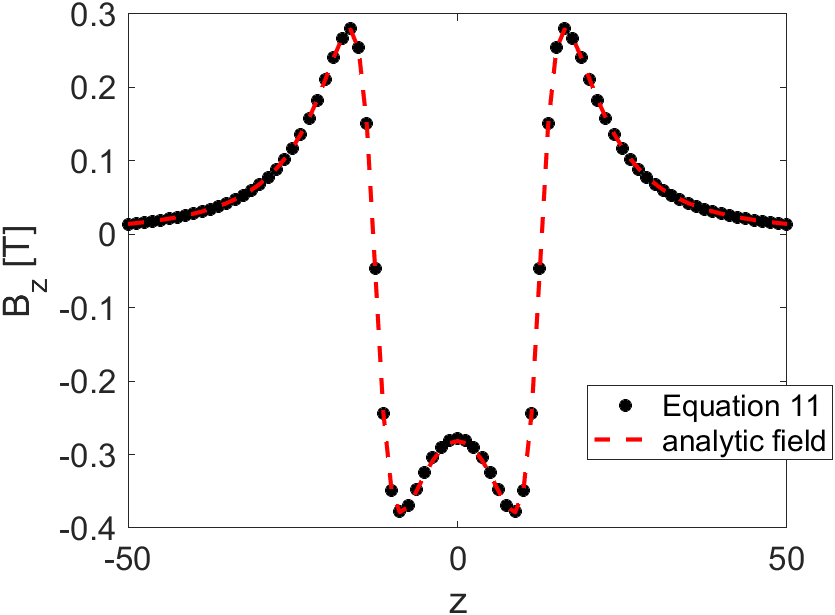}
\end{center}
\caption{\label{fig:sol} The geometry of a permanent magnet solenoid (left) and
  the field on axis (right).}
\end{figure}
As last example we consider an axial permanent magnet solenoid with $B_r=1.47\,$T,
inner radius 3\,mm, and outer radius 10\,mm. It is modeled by circular currents flowing
in one direction at the inner radius and in the opposite direction at the outer radius.
The left-hand
image in Figure~\ref{fig:sol} shows the assembly, here the circular magnet is approximated
by a hollow polygon with 18 sides. The field along the central axis of two long concentric
circular solenoids is known analytically~\cite{PENG}. We therefore compare it to our
numerical solution, based on Equation~\ref{eq:Br2}, and show both on the right-hand plot
in Figure~\ref{fig:sol}. We observe that the agreement is very good.
%
%
%
\section{Conclusions}
We determined an analytic expression for the magnetic field generated by a rectangular
current sheet and used it to model permanent magnets in iron-free geometries. We
illustrated the method by calculating the fields of simple geometries such as a single
sheet, cubes, and solenoids assembled from multiple sheets.
\par
We expect this algorithm to be useful for rapid prototyping of three-dimensio\-nal
permanent-magnet structures. It is rather fast because all integrations are done
analytically and only fields at the
points of interest need to be calculated. This is in contrast to finite-element
methods. They first need to determine the fields on a grid in the whole geometry
and then interpolate the fields to the points of interest.
On the other hand is the present method limited to iron-free geometries and to magnets
that can be modeled with rectangular current sheets.
\par
Discussions with Roger Ruber are gratefully acknowledged. In part, funding is
provided through the project {\em Disseminating technology for cold magnets to
  provide access to a wider international market} that is supported by the
European Regional Development Fund (ERDF) and Region Kronoberg.
\bibliographystyle{plain}

%
%
\appendix
\section{The integral}
\label{sec:appA}
In this appendix we evaluate the integral, given in Equation~\ref{eq:intds}. First we
consider the cross product before the integral
\begin{eqnarray}
&&  (\vr_b-\vr_a)\times(\vr_2-\vr_a-s\vec p)\nonumber\\
  &&\qquad\qquad= (\vr_b-\vr_a)\times(\vr_2-\vr_a) - \alpha s (\vr_b-\vr_a)\times\left[ (\vr_b-\vr_a)\times \vec n\right]
      \nonumber\\
  &&\qquad\qquad = (\vr_b-\vr_a)\times(\vr_2-\vr_a) + s \alpha C\vec n
\end{eqnarray}
where we used the identity $\vec a \times [\vec b \times \vec c] = \vec b (\vec a\cdot \vec c)
- \vec c (\vec a\cdot\vec b)$ to evaluate the double cross product in the second line.
Next, we calculate
\begin{equation}
A(s) = (\vr_2-\vr_a)^2-2s (\vr_2-\vr_a)\cdot \vec p +s^2 \vec p^2 = A-2D s + s^2\vec p^2
\end{equation}
where we introduce $D=(\vr_2-\vr_a)\cdot \vec p$. For $\vec p^2$ in the last term we find
\begin{eqnarray}
  \vec p^2 &=& \alpha^2 \left[(\vr_b-\vr_a)\times \vec n\right]\cdot \left[(\vr_b-\vr_a)\times \vec n\right]
  \nonumber\\
  &=& \alpha^2\left[ (\vr_b-\vr_a)^2 \vec n^2 - \left((\vr_b-\vr_a)\cdot \vec n\right)^2\right]
  =\alpha^2C\ ,
\end{eqnarray}
which allows us to write
\begin{equation}
  A(s)= A-2Ds +\alpha^2C s^2
\end{equation}
as a second-order polynomial in the variable $s$. For $B(s)$ we arrive at
\begin{eqnarray}
  B(s)
  &=& (\vr_2-\vr_a-s\vec p)\cdot (\vr_b-\vr_a)\\
  &=& B -s\alpha \left[(\vr_b-\vr_a)\times \vec n\right]\cdot(\vr_b-\vr_a) = B\ ,\nonumber
\end{eqnarray}
which is constant. Inserting these parameters into Equation~\ref{eq:intds} 
leads to
\begin{eqnarray}
 \vec B(\vr_2)
  &=&\frac{\mu_0 I'\alpha\sqrt{C}}{4\pi} \int_{-1/2}^{1/2} ds
      \frac{ (\vr_b-\vr_a)\times(\vr_2-\vr_a) + \alpha s C\vec n}{AC-B^2+C^2-2sCD+\alpha^2C^2s}\\
  &&\left[\frac{\sqrt{C}(C-B)}{\sqrt{AC-2BC+C^2-2sCD+\alpha^2 C^2 s^2}}
     +\frac{\sqrt{C}(B)}{\sqrt{AC-2sCD+\alpha^2 C^2 s^2}}\right]\ .\nonumber
\end{eqnarray}
Close inspection of the integrand shows that the second-order polynomial in the
denominator of the first line only differs by a constant term from the polynomials
under the roots in the second line. Introducing the abbreviations
\begin{equation}\label{eq:R1}
  R_1=a_1+b_1s+c_1 s^2\quad \mathrm{with}\quad a_1=AC-2BC+C^2,\ b_1=-2CD,\ c_1=\alpha^2C^2
\end{equation}
and
\begin{equation}\label{eq:R2}
R_2=a_2+b_2s+c_2 s^2\quad \mathrm{with}\quad a_2=AC,\ b_2=-2CD=b_1,\ c_2=\alpha^2C^2=c_1
\end{equation}
we can write
\begin{eqnarray}\label{eq:Blong}
  \vec B(\vr_2)
  &=&\frac{\mu_0 I'\alpha C}{4\pi} \int_{-1/2}^{1/2} ds\left[
      \frac{ (\vr_b-\vr_a)\times(\vr_2-\vr_a) + \alpha s C\vec n}{p_1+R_1}
      \frac{C-B}{\sqrt{R_1}}\right.\nonumber\\
&& \qquad\qquad\qquad\quad \left. + \frac{ (\vr_b-\vr_a)\times(\vr_2-\vr_a) + \alpha s C\vec n}{p_2+R_2}
   \frac{B}{\sqrt{R_2}}\right]\nonumber\\
  &=& \frac{\mu_0 I'\alpha C}{4\pi}\left[ (\vr_b-\vr_a)\times(\vr_2-\vr_a) (C-B)
      \int_{-1/2}^{1/2} \frac{ds}{(p_1+R_1)\sqrt{R_1}}\right.\nonumber\\
  && \qquad + \vec n\alpha C  (C-B)  \int_{-1/2}^{1/2} \frac{s ds}{(p_1+R_1)\sqrt{R_1}}\nonumber\\
  && \qquad + (\vr_b-\vr_a)\times(\vr_2-\vr_a) B  \int_{-1/2}^{1/2} \frac{ds}{(p_2+R_2)\sqrt{R_2}}\nonumber\\
  && \qquad\left. + \vec n\alpha C B  \int_{-1/2}^{1/2} \frac{s ds}{(p_2+R_2)\sqrt{R_2}}\right]
\end{eqnarray}
with $p_1=-(B-C)^2$ and $p_2=-B^2.$ We denote the integrals by
\begin{eqnarray}
  J_1&=& \int_{-1/2}^{1/2} \frac{ds}{(p_1+R_1)\sqrt{R_1}}\ ,
  \qquad J_2= \int_{-1/2}^{1/2} \frac{sds}{(p_1+R_1)\sqrt{R_1}}\ ,\nonumber\\
   J_3&=& \int_{-1/2}^{1/2} \frac{ds}{(p_2+R_2)\sqrt{R_2}}\ ,
  \qquad J_4= \int_{-1/2}^{1/2} \frac{sds}{(p_2+R_2)\sqrt{R_2}}\ .
\end{eqnarray}
We note that $J_3$ and $J_4$ resemble $J_1$ and $J_2$, only the constant parameters
$a_i,b_i,c_i$, and $p_i$ for $i=1,2$ differ. We therefore only consider $J_1$ and 
$J_2$ in the following. As a matter of fact, closed expressions for these integrals 
are given as number 2.284 in~\cite{GR}. In particular $J_1$ and  $J_2$ can be
determined from
\begin{equation}\label{eq:JJ}
  U J_2+V J_1=
  \int\frac{U s +V}{(p_1+R_1)\sqrt{R_1}}ds = \frac{U}{c_1} I_1(s)
    - \frac{2Vc_1-U b_1}{\sqrt{c_1^2p_1\left[b_1^2-4(a_1+p_1)c_1\right]}}I_2(s)
\end{equation}
with $I_1(s)$ given by~\cite{GR}
\begin{equation}\label{eq:I1}
  I_1(s)=\left\{
    \begin{array}{ll}
      \frac{1}{\sqrt{p_1}}\arctan\left(\sqrt{\frac{R_1}{p_1}}\right)
      & \quad\mathrm{for}\ p_1>0\\
      \frac{1}{2\sqrt{-p_1}}\ln\left(\frac{\sqrt{-p_1}-\sqrt{R_1}}{\sqrt{-p_1}+\sqrt{R_1}}\right)
      & \quad\mathrm{for}\ p_1<0\ .
    \end{array}
  \right.
\end{equation}
With $d_1=b_1^2-4(a_1+p_1)c_1$ the second contribution $I_2(s)$ is given by~\cite{GR}
\begin{equation}\label{eq:I2}
  I_2(s)=\left\{
    \begin{array}{ll}
      -\arctan\left(\sqrt\frac{p_1}{d_1}\right)\frac{b_1+2c_1 s}{\sqrt{R_1}}
      & \quad\mathrm{for}\ p_1d_1>0\\
      \frac{1}{2i}\ln\left(\frac{\sqrt{-d_1}\sqrt{R_1} + \sqrt{p_1}(b_1+2c_1s)}
      {\sqrt{-d_1}\sqrt{R_1} - \sqrt{p_1}(b_1+2c_1s)}\right)
      & \quad\mathrm{for}\ p_1d_1<0\ \mathrm{and}\ p_1>0\\
      \frac{1}{2i}\ln\left(\frac{\sqrt{d_1}\sqrt{R_1} + \sqrt{-p_1}(b_1+2c_1s)}
      {\sqrt{d_1}\sqrt{R_1} - \sqrt{-p_1}(b_1+2c_1s)}\right)
        & \quad\mathrm{for}\ p_1d_1<0\ \mathrm{and}\ p_1<0\\
    \end{array}
  \right.       
\end{equation}
where $R_1$ is a function of $s$, given in Equation~\ref{eq:R1}. We point out that
the sign of $I_2(s)$ in Equation~\ref{eq:JJ} is reversed with respect to~\cite{GR}
to make the numerically evaluated integral consistent with the value given in
Equation~\ref{eq:I2}. Inserting $J_1$ through $J_4$ in Equation~\ref{eq:Blong} and
reordering terms then leads to the result stated in Equation~\ref{eq:Br2} in the
main body of this report.
%
%
\section{MATLAB functions}
\label{sec:appB}
The following MATLAB functions implement the functionality described in
Section~\ref{sec:software} and are used to generate the plots shown in the
main body of this report. Note that {\tt sheets} is a $M\times 11$ array that
describes $M$ sheets.
\begin{itemize}
\item
  {\tt B=Bsheets(sheets,r2)}: returns $\vec B(\vr_2)$ for the sheets defined by
  the input argument.
\item
  {\tt sheets=make\_cubez(a,alpha,Br)}: returns the four sheets for a square block with
  width $a$ and height $\alpha a$ with the easy axis pointing in the $z$-direction. 
\item
  {\tt sheets=make\_brick(a,b,c,Br)}: returns the four sheets for a rectangular block with
  sizes $a, b,$ and $c$ in the $x,y$, and $z$ direction, respectively. The easy axis
  points in the $z$-direction. 
\item
  {\tt sheets=make\_polygon(n,r,h,Br)}: returns the $n$ sheets for a polygon with $n$
  sides, where $r$ specifies the distance from the center to the corners and $h$ is
  the height in the $z$ direction. The easy axis also points in the $z$-direction. 
\item
  {\tt sheets=make\_polygon\_hollow(n,ri,ro,h,Br)}: returns the $2n$ sheets for a hollow
  polygon with $n$ sides, where $r_i$ specifies the distance from the center to
  the corners of the inner polygon and $r_o$ to the outer polygon; $h$ is the
  height in the $z$ direction. The easy axis also points in the $z$-direction.
  This function creates ``solenoidal'' permanent magnets with single call.
\item
  {\tt sheets=sheets\_translate(sheets,a)}: returns the sheets provided as input after
  adding a vector $\vec a$ to $\vr_a$ and $\vr_b$ of the sheets.
\item
  {\tt sheets=sheets\_rotate\_x(sheets,theta)}: returns the sheets provided as input
  after the corresponding vectors $\vr_a, \vr_b$, and $\vec n$ of the sheets
  are rotated by the angle $\theta$. There are corresponding functions for
  rotations around the $y$ and 
$z$-axis.
\item
  {\tt draw\_sheet(sheet)}: draws one sheet as a blue surface and adds a green
  line to indicate the direction of the current flow. A yellow arrow indicates
  the easy axis.
\item
  {\tt B=field\_along\_line(sheets,line)}: returns an array of $\vec B(\vr_2)$ 
  at the sequence of points $\vr_2$ provided in the array {\tt line}.
\end{itemize}
The use of these functions is best illustrated in the code to prepare a Halbach
multipole. The code below first prepares a 5\,mm cube and the rotates it such
that the easy axis points in the vertical direction. After defining the number
of cubes and the multipolarity we loop over the cubes. Inside the loop we first
rotate the cube as needed to generate the desired multipolarity $m$, then translate
it radially along the $x$ axis by 8\,mm before rotating into place a second time.
Last, we add the sheets of the currently handled cube to the {\tt sheets} which,
after the loop  completes, contains all the sheets that define the multipole.
\begin{verbatim}
  cube=make_cubez(5,1,Br);
  cubey=sheets_rotate_x(cube,-90);
  MM=8; % number of cubes
  m=1;  % multipolarity (m=1: dipole, 2: quadrupole)
  for k=0:MM-1
    tmp=sheets_rotate_z(cubey,-k*m*360/MM); 
    tmp=sheets_translate(tmp,[8;0;0]);
    tmp=sheets_rotate_z(tmp,-k*360/MM);
    if k==0
      sheets=tmp;
    else
      sheets=[sheets;tmp];
    end
  end
\end{verbatim}
%
%
\section{Determining the multipolarity}
\label{sec:appC}
On magnet test benches the multipolarity of magnets is often determined from
the voltage that is induced in a coil that rotates inside the magnet. In a
dipole the polarity changes once per revolution and the voltage shows a
spectral component at the revolution frequency. In a quadrupole, the polarity
changes twice per revolution and the spectrum shows a peak at the second
revolution harmonic, whose amplitude is proportional to the quadrupole
gradient.
\par
Here, we use a simplified version of this procedure and calculate the
projection of the magnetic field onto a (normalized) vector that is tangent
to a circle with radius $r$ inside the magnet. This is illustrated by the cyan
circle and tangent vector visible on the top-left image in Figure~\ref{fig:hdip}.
In much the same way as on a measurement bench, the number of polarity changes
when traveling around the circle is related to the multipolarity $m$ of the
magnet. The amplitude $\hat B$ of the Fourier harmonic directly gives the
magnetic field harmonic on the circle. For example, in a quadrupole, $\hat B$
is related to the gradient $g$ by $g=\hat B/r$.
\par
Moreover, if we denote the real and imaginary part of the largest Fourier
harmonic by $a_m+ib_m$, the roll angle $\hat\phi$ of the magnet around the
$z$-axis (pointing upwards on the top-left image in Figure~\ref{fig:hdip})
is given by $\hat\phi=(1/m)\arctan(b_m/a_m)$. In the legend of the images in
the bottom row this value is also reported. It is, however, very small, because
the magnet was not rotated.
\par
If we use the field on the cyan circle, we measure the harmonic at a given
$z$-position, similar to what a short rotating coil would do. If, on the
other hand, we integrate the field along a line extending in the $z$-direction
and perpendicular to positions on the circle, we determine the harmonics of
the integrated field, similar to what a long coil would do. This information
is reported on the bottom left image in  Figure~\ref{fig:hdip}.
%
%
%
\end{document}